\documentclass[12pt]{article}

\begin{document}
\title{Scaling properties of the cosmic background 
plasma and radiation}
\author{ A. Bershadskii}

\maketitle
\begin{center}
{ICAR, P.O. Box 31155, Jerusalem 91000, Israel}
\end{center}

\begin{abstract}
Scaling properties of the cosmic microwave background 
(CMB) radiation are studied using satellite 
(COBE-DMR maps), balloon-borne and ground-based 
(combined QMASK map) data. Quantitative consistency 
is found between the multiscaling 
properties of the COBE-DMR and QMASK CMB maps.  
Surprisingly, it is found that the observed CMB 
temperature multiscaling resembles quantitatively 
the multiscaling properties 
of fluid turbulence, that indicates primordial 
plasma turbulence as an origin of the CMB temperature 
space anisotropy.

\end{abstract}

PACS numbers: 98.70.Vc, 98.80Es, 95.30Qd

\newpage  

\section{Introduction}
In the last few years the interest in nonlinear 
(turbulent) processes in primordial plasma has been 
renewed in relation to the origin of magnetic fields 
observed in galaxies (see, for instance 
\cite{beo}-\cite{kmk} and references therein). 
It has been suggested that 
primordial magnetic fields might arise during the
early cosmic phase transitions. The plasma of 
the early universe has a high conductivity so that a primordial 
magnetic field would be imprinted on the co-moving plasma 
and would dissipate very slowly. Such a field could then 
contribute to the seed needed to understand the presently 
observed galactic magnetic fields, 
which have been measured in both the Milky Way and other 
spiral galaxies, including their halos. At earliest times, 
the magnetic fields are generated by 
particle physics processes, with length scales typical 
of particle physics. If the inflation hypothesis is 
correct, long correlation lengths can be expected following 
the inflation. It is shown in \cite{beo}
that  turbulence with its 
cascade processes is operative, and hence the scale of 
magnetic fields is considerably larger than would be the 
case if turbulence were ignored. The turbulent nature 
of the magnetic field may have interesting effects on 
various phase transitions in the early universe. 
Also, the inherent shift of energy from small to large
scales may be of interest in connection with density 
fluctuations due to the magnetic energy. In particular, 
it is shown in \cite{sb},\cite{sb2} that rotational velocity 
perturbations, induced by a tangled magnetic field can 
produce significant angular scale anisotropies in cosmic 
microwave background (CMB) radiation through the Doppler 
effect. The conclusions are relevant to arcminute scales \cite{sb} 
and to $l > 1000$ \cite{sb2}. These scales will be available 
for observation with the new MAP and Planck space missions. 
On the other hand, the authors of \cite{dfk} and \cite{kmk} 
consider early universe turbulence and its imprint on the cosmic 
background on larger angular scales via a tensor mode contribution. 
The question then is: Can one use the modern data on the 
CMB radiation (see for recent reviews \cite{gs2},\cite{hd2}) to find 
out the fingerprints of the primordial turbulence?  The first 
attempt to find such fingerprints was performed in 
\cite{ber1}. However, a CMB map used in Ref. \cite{ber1} was one of 
the earliest of the COBE-satellite CMB maps and, therefore, the 
preliminary investigation  of  Ref. \cite{ber1}  gave a qualitative 
picture only. In present paper to obtain quantitative results we will use 
the improved  four-year COBE-satellite CMB maps. Moreover, we have 
now a new generation 
of balloon-borne and ground-based CMB data obtained with angular 
resolution of the order of a degree (the angle resolution for 
COBE data, in contrast, was $\sim 7^{0}$). We choose so-called 
QMASK map representing the data in a form suitable for our purpose. 
The QMASK map is a combination of QMAP (balloon-borne) and 
Saskatoon (ground-based) data \cite{xto}-\cite{xto2}. Using the 
three CMB maps (two COBE-DMR and one QMASK) we study the 
structure functions of the CMB temperature space increments 
and moments of CMB dissipation rate for frequency 
ranges:  31.5,53, and 90 GHz (COBE-DMR data), 26-46 GHz (QMASK map). \\

 The main results of our investigation are:

1) The CMB temperature structure functions and 
dissipation rate moments are found exhibiting multiscaling 
properties.

2) The observed multiscaling resembles quantitatively the multiscaling 
properties of fluid turbulence, that indicates its primordial 
turbulence origin. 

3) Comparison of the results obtained for the COBE-DMR and 
for the QMASK shows quantitative consistency of their CMB maps.

\section{Structure functions of QMASK map}

Let us start from the QMASK CMB map. The QMAP (balloon) and 
Saskatoon (ground-based) data were used to construct this map 
\cite{xto}-\cite{xto2}. The observations were made in Ka and Q bands 
(26-36 GHz and 36-46 GHz respectively) for different 
polarization channels. QMAP was designed to measure the 
CMB anisotropy by direct mapping. The Saskatoon data set 
is different in the sense that it does not contain 
simple sky temperature measurements. Instead, it 
contains different linear combinations
of sky temperatures with complex set of weighting 
functions. The QMAP and Saskatoon data were combined to 
produce a CMB map named QMASK. The map was generated 
by subdividing the sky into 
square pixels of side $\Theta \simeq 0.3^{o}$ and 
consists of 6495 pixels. 
The data are represented using three coordinates 
$x,y,z$ of a unit vector ${\bf R}$ in the direction of the pixel 
in the map (in equatorial coordinates). Since some 
pixels are much noisier than
others and there are noise correlations between pixels, 
Wiener filtered maps are more useful than the raw ones. 
Wiener filtering suppresses the noisiest modes in a map 
and shows the signal that is statistically significant. \\

The main tool for extracting cosmological information 
from the CMB maps is the angular power spectra. However, 
such spectra use only a fraction of the information on 
hand. Taking advantage of the good 
angular resolution of the QMASK map, we can
study statistical properties of angular {\it increments} 
of CMB temperature for different values of angular 
separation \cite{bs}. The increments are defined as
$$
\Delta T_r = (T({\bf R}+{\bf r}) - T({\bf R}))   
\eqno{(1)}
$$
where ${\bf r}$ is dimensionless vector connecting two 
pixels of the map separated by a distance $r$, and the 
structure functions of order
$p$ as $\langle|\Delta T_r|^p \rangle$ where $\langle.
\rangle$ means a statistical average over the map.

  Figure 1 shows, in log-log scales, the dependence of 
the moments of different orders, $p=0.2,0.5,0.7,1,2,3$  
of $\langle|\Delta T_r|^p \rangle$ against $r$ for the 
Wiener-filtered QMASK map. The best
linear fits are drawn to indicate the scaling
$$
\langle|\Delta T_r|^p \rangle \sim r^{\zeta_p}.  
\eqno{(2)}
$$
Figure 2 shows the scaling exponents $\zeta_p$ extracted 
from the QMASK data shown in Fig. 1 (circles). 
We also show in this figure 
(as dashed curve) the scaling exponents $\zeta_p$ 
corresponding to velocity increments in fluid turbulence.  
For this purpose we use the She-Leveque model \cite{sl}, 
which is in very good agreement with the numerous data 
for velocity increments obtained in inertial interval of 
scales for isotropic fluid turbulence:
$$
\zeta_p = p/9 + 2[1-(2/3)^{p/3}]      \eqno{(3)}
$$
The correspondence seen in Figure 2 is rather remarkable 
(especially for low moments for which the
calculations are also the most reliable).  

It appears that for the angular interval between 
$0.9^{o}$ and $4^{o}$---this being the most reliable 
range for the data \cite{xto2}---the temperature 
increments in QMASK map, scale essentially as velocity 
increments in fluid turbulence.
 
 It should be noted that under certain circumstances 
scaling of the moments (2) for $p=2$ can be related 
to scaling of corresponding spectrum. One of the 
necessary conditions for such 
relation is sufficiently fast decrease of the spectrum 
with increasing of wave numbers $k > \eta^{-1}$, where 
$\eta$ is the lower edge of the scaling interval of $r$. 
For fluid turbulence such fast decay of the spectral 
asymptotic indeed takes place ($\eta$ is so-called 
Kolmogorov or {\it viscous} scale there). However, 
it seems that for the CMB data this is not the case 
and the strong relation between fluid turbulence and 
the CMB space fluctuations takes place in a 
restricted interval of scales only (and does not 
determine the spectral asymptotic).     \\ 

 Background signal can be distinguished from the 
atmospheric and foreground 
contamination signals by their frequency dependence, 
their frequency coherence 
and their spatial power spectra. Employing these 
methods, it has been shown in a series of papers 
\cite{xto}-\cite{xto2},\cite{max},\cite{miller},\cite{oliv2} 
that all known atmospheric and foreground sources 
give a minor contribution to the QMASK map. 
We are thus led to conclude that the background 
(primordial plasma) turbulence is the most plausible reason 
why the CMB map resembles fluid turbulence in its intermittency 
properties. It should be noted that remarkable correspondence 
between the scaling laws in the {\it near} space plasma and in 
isotropic fluid turbulence is a well known puzzle (see, for 
instance, \cite{gold},\cite{clv} and references therein).   

\section{Dissipation rate of CMB temperature from 
the four-year COBE-DMR data maps} 

  The above discovered modulation of the CMB temperature 
by a fluid turbulent velocity fluctuations can be 
explored further. Dissipation rate of kinetic energy 
corresponding to the velocity field ${\bf u}$ is \cite{my}:  
$$
\varepsilon \sim \sum_{ij} (\frac{\partial u_i}{\partial 
x_j})^2  \eqno{(4)}
$$ 
Local space averaging of the dissipation rate is used 
to construct a space dissipation measure \cite{my}:
$$
\varepsilon_r =\frac{\int_{v_r} \varepsilon dv}{v_r}
    \eqno{(5)}
$$
where $v_r$ is a subvolume with space-scale $r$. 
Scaling law of this measure moments,
$$
\langle \varepsilon_{r}^p \rangle \sim r^{-\mu_p}      
\eqno{(6)}
$$
(where $\langle ... \rangle$ means an average) is an 
important characteristic of the dissipation rate 
\cite{my}. For isotropic fluid turbulence there 
exists a simple 
relation between scaling exponents $\zeta_p$ for 
the velocity increments (see previous section) and 
the scaling exponents $\mu_p$ (6)  \cite{my}: 
$$
\mu_p=p -\zeta_{3p}    \eqno{(7)}
$$
This relationship allows to calculate $\mu_p$ 
using the She-Leveque representation (3). 

  Let us now calculate the dissipation rate of the 
CMB temperature fluctuations using the COBE-DMR CMB 
maps. The COBE satellite was launched on 1989 into 900 km 
altitude sun-synchronous orbit. The Differential Microwave 
Radiometers (DMR) operated for four years of 
the COBE mission and mapped the full sky. 
The instrument consists of six differential microwave 
radiometers, two nearly independent channels that operate at each of 
three frequencies: 31.5, 53 and 90 GHz. 
Each differential radiometer 
measures the difference in power received from two 
directions in the 
sky separated by 60$^0$, using a pair of horn antennas. 
Each antenna has a 7$^0$ beam. In this investigation we 
will use a four-year DMR data map of the CMB temperature. 
Most of the Galactic emission has been removed from the map 
using so-called "combination technique" in which 
a linear combination of the DMR maps is used to 
cancel the Galactic emission \cite{benn1}. 
We will also use 
a DMR four-year sky map for frequency 31.5 GHz only. 
In the last case we will remove the sky regions with 
violently large temperature fluctuations to remove 
roughly the Galactic contamination (see 
below and Fig. 4). The first detail interpretations of the four-year 
COBE/DMR data one can find in \cite{benn2}-\cite{fix}.

To organize the DMR data 
in the temperature maps the sky was divided into 
6144 equal area pixels. These 
are formed by constructing a cube with each face 
divided into 32 x 32 = 1024 
squares, projected onto a celestial sphere in 
elliptic coordinates. The projection 
is adjusted to form equal area pixels having a 
solid angle of 4$\pi$/6144 sr or 6.7 
square degrees. Because the $7^0$ beamwidth of 
the sky horn is greater than 
separation between pixels ($2.6^0$ average), 
this binning oversamples the sky. \\

Let us introduce a dissipation rate for CMB 
temperature $T$ in following way (cf (4),(5)):
$$
\chi_r =\frac{\int_{v_r} (\bigtriangledown{T})^2 dv}{v_r}    
\eqno{(8)}
$$

  Technically, using the above described pixel map, 
we will calculate the CMB temperature dissipation rate 
using summation over pixel sets instead of integration over 
subvolumes $v_r$. So that in the multiscaling 
$$
\langle \chi_s^p \rangle \sim s^{-\mu_p}      
\eqno{(9)}
$$
the metric scale $r$ is replaced by 
number of the pixels - $s$, characterizing size of the 
summation set (cf (6)). The $\chi_s$ is a surrogate 
of the real 3D dissipation rate $\chi_r$. 
It is believed that the surrogates can 
reproduce quantitative multiscaling properties of 
the 3D dissipation rates \cite{sa}.     

Figure 3 shows scaling of the CMB temperature 
dissipation rate moments $\langle \chi_s^p \rangle$ 
calculated for the DMR maps (Fig 3a: for combination of 
frequencies 31.5, 53 and 90 GHz; Fig. 3b: for 
frequency 31.5 GHz). Log-log scales are chosen in 
the figure for comparison with scaling equation (9). 
The straight lines (the best fit) are drawn to indicate 
the scaling. 

Figure 4 shows the scaling exponents $\mu_p$  
extracted from figure 3a (open circles) and from 
figure 3b (crosses), the dashed curve in figure 4 
corresponds to the intermittency exponents $\mu_p$ 
calculated using the She-Leveque model \cite{sl} for 
velocity (kinetic energy) fluctuations of fluid 
turbulence: equations (3) and (7). 

   Thus, for the satellite data we also 
observe good agreement with fluid turbulence 
modulation of the CMB temperature. 
Applicability of the fluid turbulence relationship 
(7) to the CMB data can be itself considered as an additional 
indication of this modulation. Moreover, 
comparing Figures 2 and 4 (through the relationship 
(7)) one can conclude that the QMASK data represented 
in Figure 2 are quantitatively consistent to the COBE-DMR 
data represented in Figure 4. In this respect it should 
be noted that the QMASK and the COBE-DMR data were 
obtained in about the same range of frequencies. 
Very recent suggestions to detect the primordial 
turbulence using the relic gravitational waves 
\cite{kmk},\cite{dgn},\cite{dg} might give a complementary 
experimental tool for the present observations in the nearest 
future.\\

The author is grateful to the QMAP, Saskatoon, COBE-DMR 
instruments teams and to the NASA Goddard Space Flight 
Center for providing the data and support. Discussions 
on the subject with C.H. Gibson, A. Kosowsky and K.R. Sreenivasan 
as well as the Referee's comments and information were useful 
for this investigation.

\newpage

\begin{center}
Figure Captions
\end{center}
\bigskip
{\noindent \parindent 0pt \parskip 5pt}

Figure 1.  Logarithm of the structure functions 
$\langle|\Delta T_r|^p \rangle$ against $log_{10} r$ for the
Wiener-filtered QMASK map. The CMB temperature $T$ 
is measured in $\mu K$. The straight lines (the best fit) 
are drawn to indicate the scaling (2). \\

Figure 2. The scaling exponents $\zeta_p$, corresponding to
Eq.~(2), for the QMASK map (circles). The dashed curve corresponds 
to the scaling exponents $\zeta_p$ calculated using the She-Leveque 
model (3) \cite{sl} for the turbulent fluid velocity increments.\\

Figure 3. a) The CMB dissipation rate moments $\langle \chi_s^p 
\rangle$ against $s$ (the COBE-DMR four-year map for combination of 
frequencies: 31.5, 53 and 90 GHz). 
The dissipation rate $\chi_s$ is measured in 
$10^{-2}~\mu K^2$; b) Analogous data, but now for the CMB sky 
four-year map corresponding to frequency 31.5 GHz.\\

Figure 4. The scaling exponents $\mu_p$ extracted from figure 3 
against $p$. Open circles correspond to the CMB four-year map 
for combination of frequencies: 31.5, 53 and 90 GHz, 
and crosses correspond 
to the CMB four-year sky map for frequency 31.5 GHz. The dashed 
curve corresponds to the intermittency exponents $\mu_p$ calculated 
using the She-Leveque model for the kinetic energy dissipation 
in fluid turbulence (equations (3),(7)) \cite{sl}. 

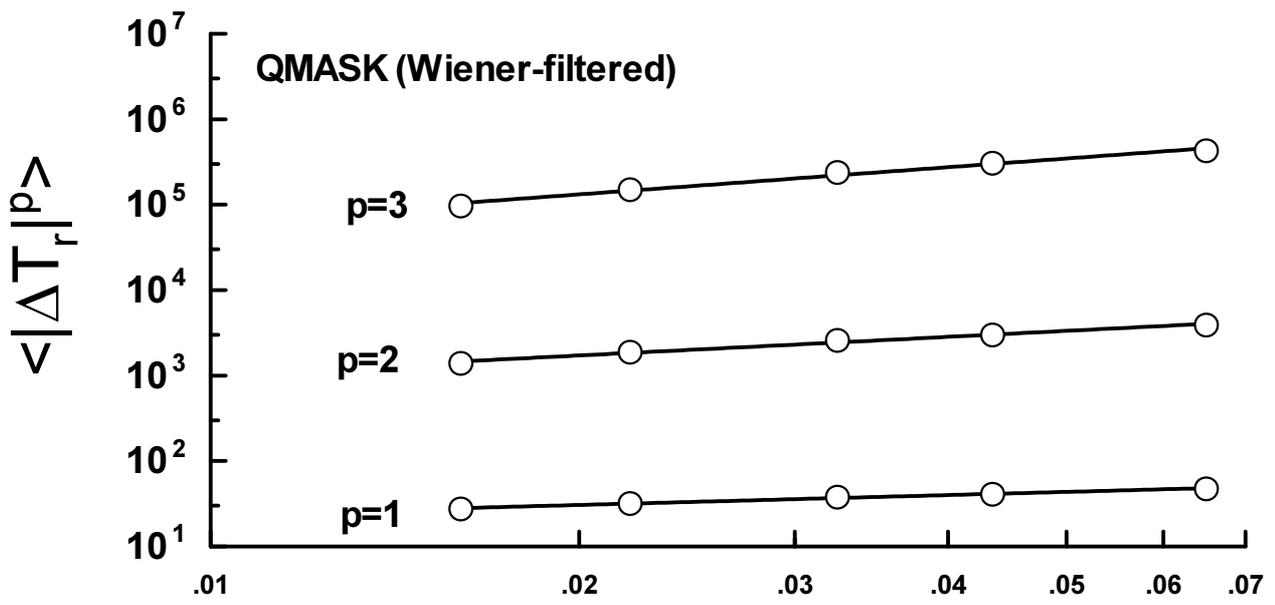
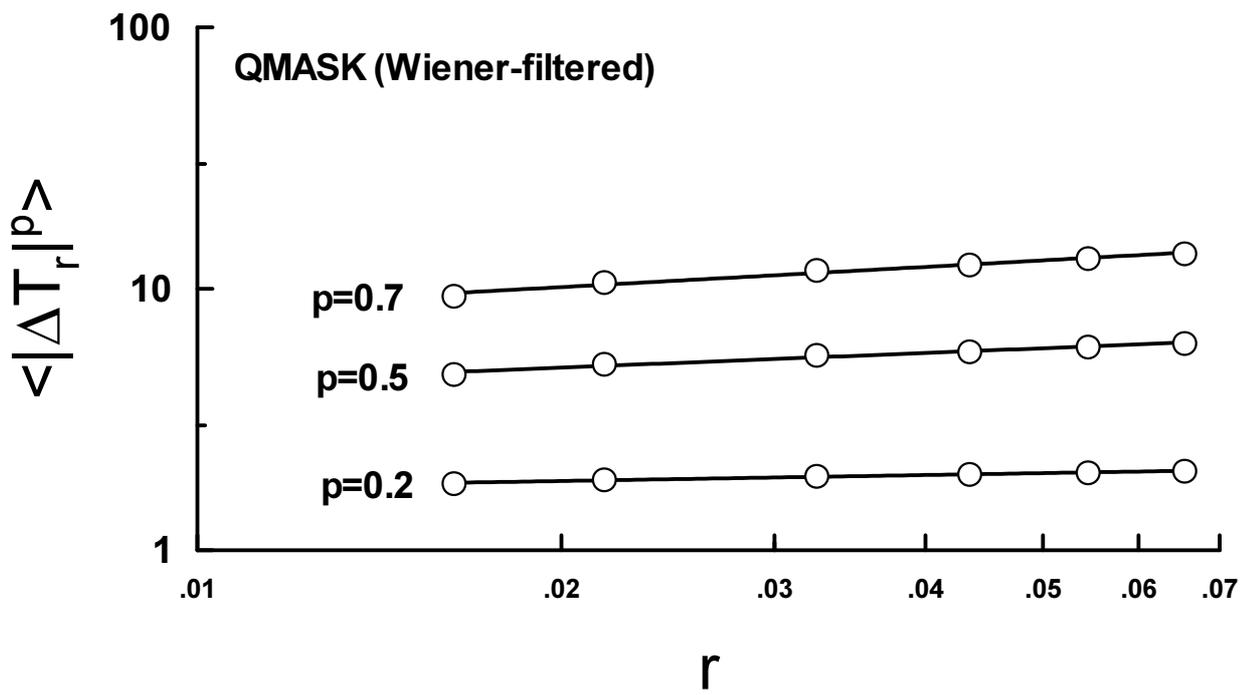
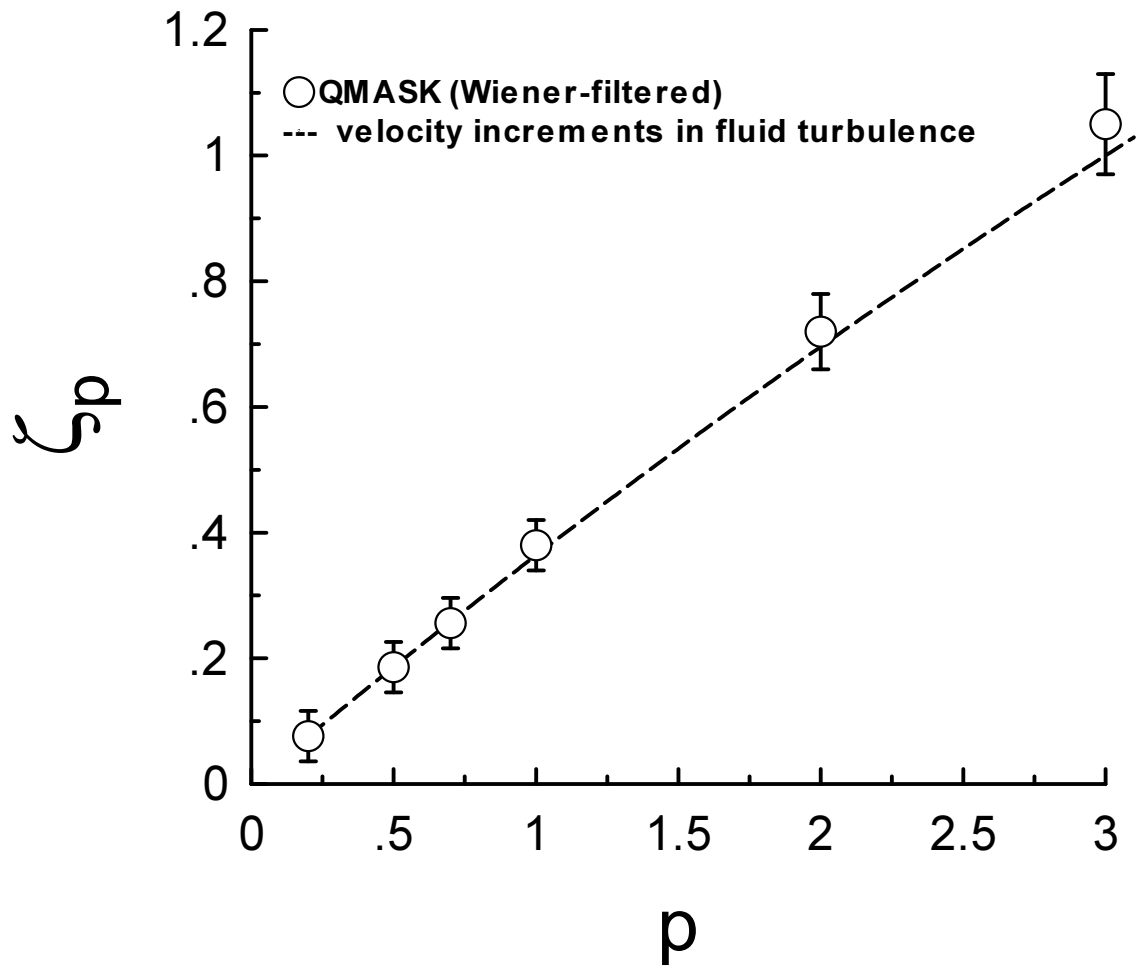
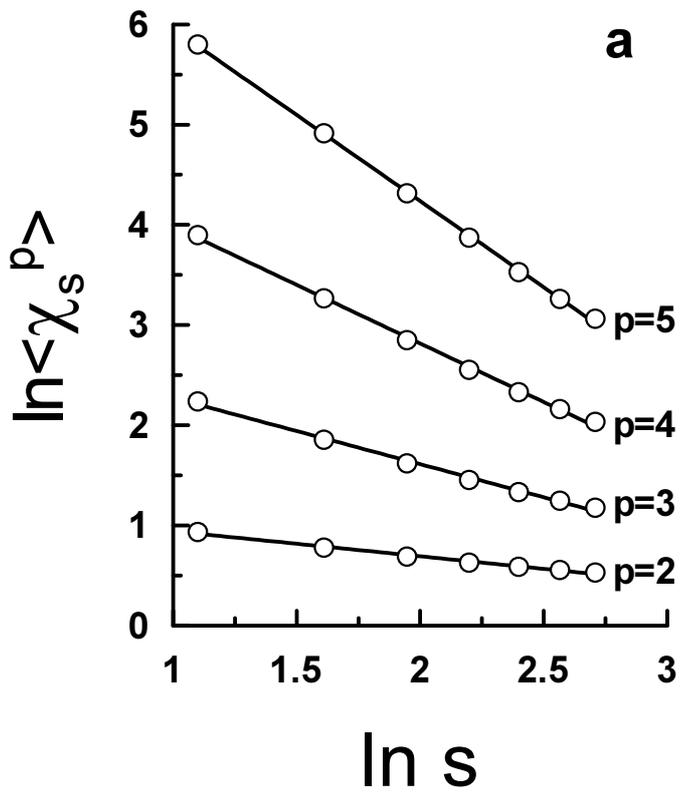
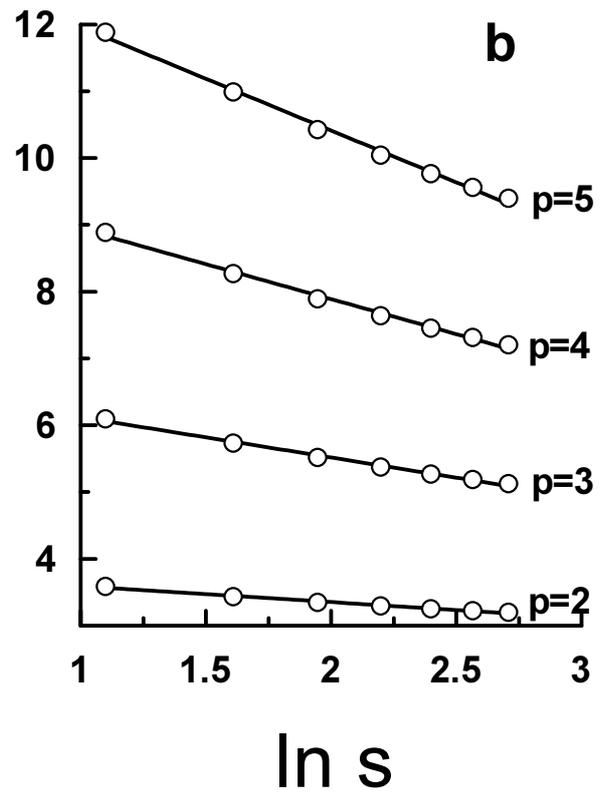
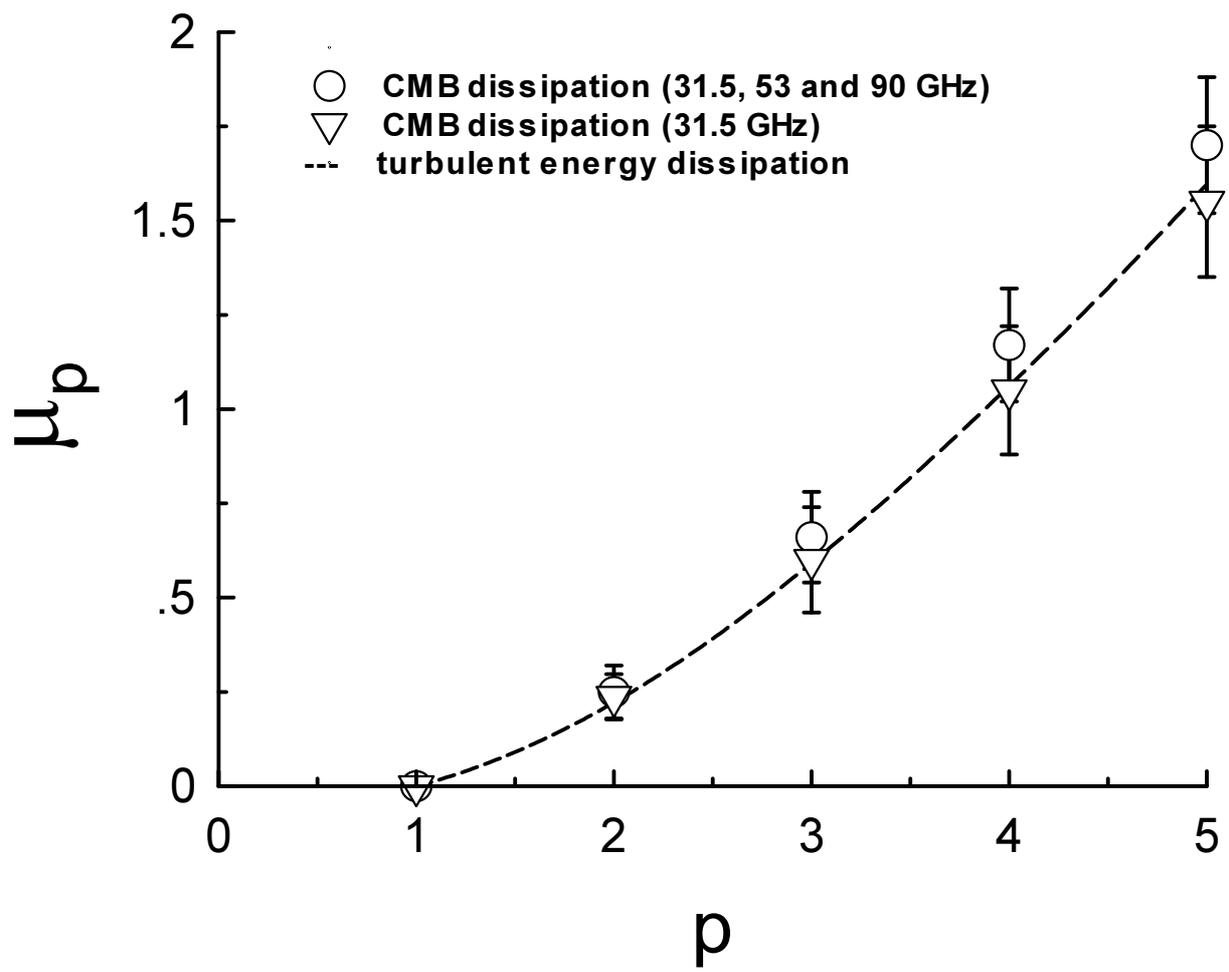
\end{document}